\documentclass[prc,aps,
twocolumn,
tightenlines,
preprintnumbers,
superscriptaddress,
a4paper,nofootinbib]{revtex4}
\usepackage{graphicx}
\usepackage{tikz}
\usepackage{epsf}
\usepackage{amsmath,amssymb}             
\usepackage{amsfonts}
\usepackage{mathrsfs}
\usepackage{xcolor}
\usepackage{cancel}
\usepackage{nicefrac}
\usepackage{array}

\usepackage{times}
\usepackage{mathptmx}   

\newcolumntype{L}[1]{>{\raggedright\let\newline\\\arraybackslash\hspace{0pt}}m{#1}}
\newcolumntype{C}[1]{>{\centering\let\newline\\\arraybackslash\hspace{0pt}}m{#1}}
\newcolumntype{R}[1]{>{\raggedleft\let\newline\\\arraybackslash\hspace{0pt}}m{#1}}

\usepackage[colorlinks,citecolor=blue,linktoc=all,linkcolor=cyan]{hyperref} 

\def\beq{\begin{equation}}
\def\eeq{\end{equation}}
\def\bea{\begin{eqnarray}}
\def\eea{\end{eqnarray}}
\def\beqa{\begin{equation}\begin{array}{l}}
\def\eeqa{\end{array}\end{equation}}
\def\eqlab#1{\label{eq:#1}}
\def\figlab#1{\label{fig:#1}}

\def\Eqref#1{Eq.~(\ref{eq:#1})}

\def\Figref#1{Fig.~\ref{fig:#1}}

\def\barr{\left(\begin{array}{c}}
\def\earr{\end{array}\right)}
\def\bmat{\left(\begin{array}{cc}}
\def\emat{\end{array}\right)}
\def\al{\alpha}
\def\ga{\gamma} 
 \def\De{\Delta}

\def\la{\lambda}

\def\si{\sigma} 
\def\th{\theta}  
\def\w{\omega}

\def\dd{{\rm d}}

\def\nn{\nonumber}

\DeclareMathOperator\re{Re}
\def\3d{3-D}

\newcommand{\mypentagon}[1]{
\begin{tikzpicture}
\fill[opacity=1,#1] (1ex, 0.5257311121191336ex) -- (0.6180339887498949ex,1.0514622242382672ex) -- (0, 0.8506508083520399ex) -- (0, 0.2008114158862273ex) -- (0.6180339887498949ex, 0) -- (1ex, 0.5257311121191336ex);
\end{tikzpicture}
}
\newcommand{\mytriangle}[1]{
\begin{tikzpicture}
\fill[opacity=1,#1] (-0.5ex, 0) -- (0.5ex, 0) -- (0.ex, 0.866025403784ex) -- (-0.5 ex,0);
\end{tikzpicture}
}
\definecolor{olive}{HTML}{668000}
\definecolor{lightolive}{HTML}{CCFF00}
\definecolor{myorange}{HTML}{FF9900}
\definecolor{mylblue}{HTML}{8080FF}
\definecolor{mydgreen}{HTML}{008000}

\begin{document}
\preprint{MITP/17-100}

\title{Partial-wave analysis of proton  
Compton scattering data below the pion-production threshold}

\author{Nadiia Krupina}
\affiliation{Institut f\"ur Kernphysik \& Cluster of Excellence PRISMA, Johannes Gutenberg Universit\"at, Mainz D-55099, Germany}
\author{Vadim Lensky}
\affiliation{Institut f\"ur Kernphysik \& Cluster of Excellence PRISMA, Johannes Gutenberg Universit\"at, Mainz D-55099, Germany}
\affiliation{Institute for Theoretical and Experimental Physics, Bol'shaya Cheremushkinskaya 25, 117218 Moscow, Russia}
\affiliation{National Research Nuclear University MEPhI (Moscow Engineering Physics Institute), 115409 Moscow, Russia}
\author{Vladimir Pascalutsa}
\affiliation{Institut f\"ur Kernphysik \& Cluster of Excellence PRISMA, Johannes Gutenberg Universit\"at, Mainz D-55099, Germany}

\date{\today}

\begin{abstract}
Low-energy Compton scattering off the proton is used for 
determination of the proton polarizabilities.  However,
the present empirical  determinations rely  heavily on 
the theoretical description(s) of the experimental cross sections
in terms of polarizabilities. The  most recent determinations
are based on either the fixed-$t$ dispersion relations (DR) or
chiral perturbation theory in the single-baryon sector ($\chi$PT). 
The two approaches obtain rather different results for 
proton polarizabilities, most notably for $\beta_{M1}$ (magnetic dipole
polarizability).
We attempt to resolve this discrepancy by performing a
partial-wave analysis of the 
world data on  proton  Compton scattering below threshold.
We find a large sensitivity of the extraction to a few  ``outliers'',
leading us to conclude that the difference between DR and $\chi$PT extraction is
a problem of the experimental database rather than  of  ``model-dependence''.
We have specific suggestions for new experiments needed for an
efficient improvement of the database. With the present database, 
the difference of proton scalar polarizabilities is constrained
to a rather broad interval:
$\alpha_{E1}-\beta_{M1} = (6.8\, \ldots \, 10.9)\times 10^{-4}$ fm$^3$, with their 
sum fixed much more precisely [to $14.0(2)$] by the Baldin sum rule. 

\end{abstract}


\maketitle
%
%

\newpage
\section{Introduction}

The low-energy Compton scattering (CS) off the proton and light 
nuclei
is the standard tool for probing the {\it nucleon polarizabilities}, see~\cite{Drechsel:2002ar,Schumacher:2005an,Griesshammer:2012we,Holstein:2013kia,Hagelstein:2015egb}
for reviews. However, the relation  between the  experimental
observables and polarizabilities is simple only when neglecting
the higher-order terms in the low-energy expansion (LEX) of Compton amplitudes. In practice, the higher-order terms 
play an important role, and, for a quantitative 
extraction of polarizabilities from Compton scattering data, more sophisticated theoretical frameworks are being used.
In the case of the proton, there are two types
of ``state-of-the-art'' extractions: (i) based on
the fixed-$t$ dispersion relations (DR) \cite{Drechsel:2002ar,Schumacher:2005an,Lvov:1993fp,Lvov:1996rmi,Drechsel:1999rf,Pasquini:2007hf}
and (ii) based on chiral perturbation theory ($\chi$PT)
with explicit nucleons and Delta's. The latter calculations
can be divided into two types:
heavy-baryon (HB$\chi$PT) \cite{Bernard:1995dp,Beane:2002wn,McGovern:2012ew} 
or manifestly-covariant
(B$\chi$PT)~\cite{Lensky:2009uv,Lensky:2015awa}.
The problem is that, although both DR and $\chi$PT give 
comparably good description of the experimental cross sections, the extracted
values of polarizabilities differ, sometimes by  a few
standard deviations.  

A notable example is  provided by the magnetic dipole polarizability $\beta_{M1}$ of the proton, which  ranges from  $1.6(4)$ [in units of $10^{-4} \;\text{fm}^3$, omitted in what follows] obtained in the state-of-art DR fits of the data~\cite{Schumacher:2005an,Pasquini:2007hf,OlmosdeLeon:2001zn} to $3.2(5)$ in the $\chi$PT fits~\cite{Griesshammer:2012we,McGovern:2012ew,Lensky:2014efa}. 
Furthermore,  without using the Compton data, the B$\chi$PT at NNLO yields for the proton~\cite{Lensky:2009uv,Lensky:2015awa}: $\beta_{M1}=3.9(7)$, 
making the discrepancy with DR more acute.  Incidentally, the current {\it PDG average}~\cite{Patrignani:2016xqp}  is basically combining  
the DR and HB$\chi$PT values, resulting in $2.5(4)$ for the proton $\beta_{M1}$.
Their central value may serve as a compromise, but the uncertainty does not seem 
to reflect the spread between the DR and $\chi$PT values.

The present work is an attempt to resolve this tension in a model-independent  manner, by making the partial-wave analysis (PWA) of the CS data below the pion photoproduction 
threshold ($\leqslant 150$ MeV).  To our surprise, we find that the above discrepancy
between DR and $\chi$PT fits is a problem of the experimental database, rather than of theoretical descriptions. 
As such, it calls for new experiments.
We argue that new precision data for the proton CS angular distribution at backward angles and beam energy around 100 MeV are highly desirable.

The PWA is of course a good old method to study the hadronic processes at low energy.  Yet, it has barely been used in 
proton CS ($\ga p\to \ga p$). 
To date, the only comprehensive PWA of proton CS data remains to be the 1974 analysis of 
Pfeil et al.~\cite{Pfeil:1974ib}, in the $\De$(1232) region. The region below the pion threshold has not been analyzed until now --- the present PWA is the first. 

Of course, this is not a first study
of the Compton multipoles below threshold in general, cf.\ 
Refs.~\cite{Hildebrandt:2003fm,Hildebrandt:2005ix,Lensky:2015awa,Pasquini:2017ehj}
 for calculations using DR and $\chi$PT frameworks. However, 
a model-independent Compton PWA has not done until now,  
mainly due to the lack of accurate data. 
The latter problem is compensated in our analysis 
by  the recent empirical determination
of the forward Compton amplitudes through the sum rules 
involving the photoabsorption cross sections \cite{Gryniuk:2015eza,Gryniuk:2016gnm}. The sum rules
yield two independent linear relations between the 
multipole amplitudes at each energy. Note that the linear relations  among the partial-wave amplitudes are very rare in PWAs. They usually operate with the bi-linear relations between the amplitudes and experimental observables alone. Furthermore, 
 we make use of the fact that
the partial-wave (or multipole) amplitudes are
assumed real below the threshold, that is if one neglects the radiative corrections. 
It is also important to build in the correct low-energy limit and treat exactly the Born contributions.

In the following we present the 
low-energy parametrization of the pertinent multipole amplitudes
in terms of the static polarizabilities  (Sec.~\ref{sec:Formalism}),
and the corresponding fits of the experimental data for proton CS (Sec.~\ref{sec:fitting}). The results are discussed in 
Sec.~\ref{sec:results}, and conclusions are given in Sec.~\ref{sec:Conclusions}. 

\setlength{\abovedisplayskip}{6pt plus 2pt minus 4pt}
\setlength{\belowdisplayskip}{6pt plus 2pt minus 4pt}

\section{Multipole expansion and (bi-)linear empirical constraints}
\label{sec:Formalism}

The general formalism of the multipole expansion for nucleon CS is given in \cite{Pfeil:1974ib,Krupina:2016}, and 
concisely summarized in \cite[Sec.~2]{Hagelstein:2015egb}.
The idea is that, using the rotational and discrete symmetries,  the CS helicity amplitudes $T_{\sigma' \lambda', \sigma \lambda}(s,t)$,
with $\sigma$ ($\sigma'$) the helicity of initial (final) photon and  $\lambda$ ($\lambda')$ for the helicity initial (final) nucleon, admits a partial-wave
expansion:
\beq
T_{\sigma' \lambda', \sigma \lambda} = \sum_{J=1/2}^{\infty} (2J+1)\, T_{\sigma' \lambda', \sigma \lambda}^J(\w) \,d_{\sigma'-\lambda',\, \sigma-\lambda}^J(\theta),
\eeq
with $J$ the total angular momentum of the photon-proton system, $T_{\sigma' \lambda', \sigma \lambda}^J(\w)$ the partial-wave amplitudes,  $d_{\lambda',\,\lambda}^J(\theta)$ the Wigner $d$-function,  $\w$ and  $\theta$ the photon energy and scattering angle in the center-of-mass   frame;  $s$, $t$, $u$ are Mandelstam invariants.

The partial-wave amplitudes $T^J(\w)$ are then linearly related to the
amplitudes with definite parity and angular momentum $l$, i.e., multipoles
 $f_{\rho\rho'}^{l \pm}(\w)$, with $\rho, \rho'= E\mbox{(lectric), or } M\mbox{(agnetic)}$. The infinite sum over half-integer $J$ is then replaced by the sum over integer
 $l = J\mp 1/2$. Note that
 $f_{\rho\rho'}^{0+}=0$, by definition; hence the
 summation starts at $l=1$.

In this work, we first write the amplitude as the sum of the Born, $T^{\mathrm{Born}}$, and  the rest (non-Born) $\bar T$, as illustrated in Fig.~\ref{fig:Born} (note that here the $\pi^0$-pole contribution is part of the Born term). The same decomposition holds for 
the multipoles: $f = f^{\mathrm{Born}} + \bar f$.
We then truncate the multipole expansion of the non-Born amplitude at 
$J=3/2$, whereas the Born amplitude is treated exactly.
We thus retain the ten lowest non-Born multipoles,
\beq 
\label{eq:ten_multipoles}
\bar{f} = \big( \bar{f}_{EE}^{1+},\; \bar{f}_{EE}^{1-},\; \bar{f}_{MM}^{1+},\; \bar{f}_{MM}^{1-},\; \bar{f}_{EM}^{1+},\; \bar{f}_{ME}^{1+},
\; \bar{f}_{EE}^{2+},\; \bar{f}_{EE}^{2-},\; \bar{f}_{MM}^{2+},\; \bar{f}_{MM}^{2-}\,\big),
\eeq 
the rest are neglected. This approximation is well justified at energies
below the pion production threshold ($\w\lesssim m_\pi$), as the leading low-energy behavior of the non-Born multipoles is~\cite{Guiasu:1978dz}
\beq
\eqlab{LObehav}
\bar{f}^{l\pm}_{EE}\sim \bar{f}^{l\pm}_{MM}\sim \omega^{2l}, \qquad \bar{f}^{l+}_{EM}\sim \bar{f}^{l+}_{ME}\sim  \omega^{2l+1}\,.
\eeq 
Furthermore, the existing $\chi$PT calculations \cite{Lensky:2009uv,Lensky:2015awa,Hildebrandt:2003fm} show that the four $l=2$ non-Born multipoles, $\bar f_{EE}^{2+}$, $\bar f_{EE}^{2-}$, $\bar f_{MM}^{2+}$, $\bar f_{MM}^{2-}$, give tiny contributions below the pion threshold. In what follows we will either
neglect them, or fix them to the values given by the latest B$\chi$PT calculation~\cite{Lensky:2015awa}. 
We shall therefore fit only the six $l=1$ non-Born multipoles.

In order to build in the low-energy behavior
of the non-Born multipoles [cf. \Eqref{LObehav}], we assume
the following parametrization of the $l=1$ multipoles in terms of static polarizabilities:
\begin{widetext}
\begin{figure*}[htb]
\includegraphics[width=0.7\linewidth]{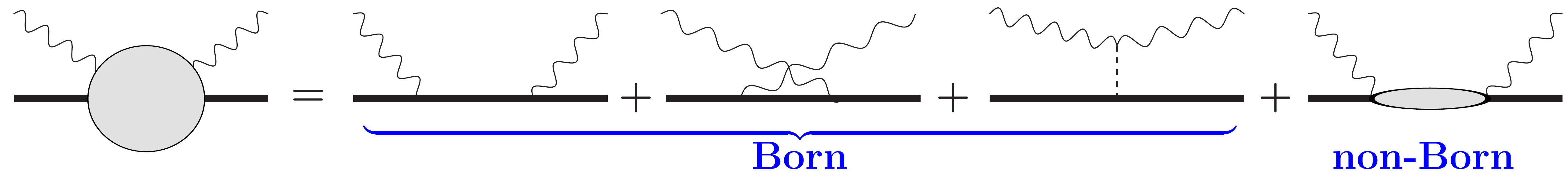}
\caption{Mechanisms contributing to real CS: Born and non-Born terms.}
\label{fig:Born}
\end{figure*}

\bea
\eqlab{nonBorn_multipoles_expansion}
&&\bar f_{EE}^{1+}(E_\gamma) = E_\gamma^2 \,\frac{M}{\sqrt{s}} \left[\frac{\alpha_{E1}}{3} + \frac{E_\gamma}{3}\,\left(\frac{-\alpha_{E1}+\beta_{M1}}{M} + \gamma_{E1E1}\right) + \left(\frac{E_\gamma}{M}\right)^2 f_1^{R}(E_\gamma)\right], \nn \\
&&\bar f_{EE}^{1-}(E_\gamma) = E_\gamma^2 \,\frac{M}{\sqrt{s}} \left[\frac{\alpha_{E1}}{3} + \frac{E_\gamma}{3}\,\left(\frac{-\alpha_{E1}+\beta_{M1}}{M} - 2\, \gamma_{E1E1}\right) + \left(\frac{E_\gamma}{M}\right)^2 f_2^{R}(E_\gamma)\right], \nn \\
&&\bar f_{MM}^{1+}(E_\gamma) = E_\gamma^2 \,\frac{M}{\sqrt{s}} \left[\frac{\beta_{M1}}{3} + \frac{E_\gamma}{3}\,\left(\frac{-\beta_{M1}+\alpha_{E1}}{M} + \gamma_{M1M1}\right) + \left(\frac{E_\gamma}{M}\right)^2 f_3^{R}(E_\gamma)\right],  \\
&& \bar f_{MM}^{1-}(E_\gamma) = E_\gamma^2 \,\frac{M}{\sqrt{s}} \left[\frac{\beta_{M1}}{3} + \frac{E_\gamma}{3}\,\left(\frac{-\beta_{M1}+\alpha_{E1}}{M} -2\,\gamma_{M1M1}\right) + \left(\frac{E_\gamma}{M}\right)^2 f_4^{R}(E_\gamma)\right], \nn \\
&& \bar f_{EM}^{1+}(E_\gamma) = E_\gamma^3 \,\frac{M}{\sqrt{s}} \left[\frac{\gamma_{E1M2}}{6} + \frac{E_\gamma}{6}\, \left(\frac{-6\, \gamma_{E1M2} + 3\,\gamma_{M1E2} + 3\,\gamma_{M1M1}}{4 M} - \frac{\beta_{M1}}{8 M^2} \right) + \left(\frac{E_\gamma}{M}\right)^2 f_5^R (E_\gamma) \right], \nn \\
&& \bar f_{ME}^{1+}(E_\gamma) = E_\gamma^3 \,\frac{M}{\sqrt{s}} \left[\frac{\gamma_{M1E2}}{6} + \frac{E_\gamma}{6}\, \left(\frac{-6\, \gamma_{M1E2} + 3\,\gamma_{E1M2} + 3\,\gamma_{E1E1}}{4 M} - \frac{\alpha_{E1}}{8 M^2} \right) + \left(\frac{E_\gamma}{M}\right)^2 f_6^R (E_\gamma) \right], \nn
\eea
\end{widetext}
where we changed the photon energy from the center-of-mass 
$\omega$ to the lab frame $E_\gamma
=\w \sqrt{s}/M$. 
The first term in each of the square brackets of \Eqref{nonBorn_multipoles_expansion} is
given by one of the six static polarizabilities, four of which, denoted by
$\gamma$'s, are 
spin-dependent. The 2nd terms are the recoil corrections (see, e.g.,
Ref.~\cite{Lensky:2015awa}). The 3rd terms are given
by the ``residual functions" $f^R_i$. 
The parametrization of \Eqref{nonBorn_multipoles_expansion} ensures the correct low-energy behavior
of these multipoles. It does not imply
any approximation: the six static polarizabilities as well as the residual
functions are free parameters, which 
will next be determined from  experimental data. 

\subsection{Bilinear relations: observables}
\label{sec:Bilinear_relations}

Any CS observable provides {\em bi-linear} relations on CS multipoles. 
This is of course the usual situation for any PWA, an experimental observable, such as cross section or asymmetry, involves the 
amplitude squared.

Take for instance the unpolarized angular distribution $\dd\sigma/\dd\Omega$, given in terms of the helicity amplitudes by
\bea
\label{eqn:Unpol_phi_i}
\frac{\dd \sigma}{\dd \varOmega} 
&=&  \frac{1}{256 \pi^2 s} \sum_{\sigma' \, \la'  \sigma  \la }
 \big|T_{\sigma' \la',  \sigma \, \la}\big|^2 .
\label{eqn:Def_dXS_unpol}
\eea
Substituting in here the multipole expansion of $T$ we obtain (for $J< 5/2$): 
\beq 
\frac{\dd \sigma}{\dd \varOmega} =  \sum_{n=0}^4
c_n \cos n\theta,
\eeq
where $c_n$ are bilinear combinations of the multipoles.
In principle, each $c_n$ can be extracted from the fit to the data,
and hence one obtains 5 bilinear relations from this observable.

Similarly, for the beam asymmetry, defined as
\beq
\Sigma_3 = \frac{\dd \sigma_{||} -\dd \sigma_{\perp}}{\dd \sigma_{||} +\dd \sigma_{\perp}} ,
\eeq
where $\sigma_{||}$ and $\sigma_{\perp}$ are the CS cross sections with linear photon-beam polarization (parallel and perpendicular to the scattering plane), we have:
\bea 
 \frac{\dd \sigma}{\dd \varOmega} \Sigma_3 
&=&  \frac{1}{128 \pi^2 s} \sum_{\sigma' \, \la'  \la }
\re ( T^\ast_{\sigma' \la', -1 \, \la} T_{\sigma' \la', 1 \, \la}) \nn\\
&\stackrel{J< 5/2}{=}&   \sin^2\th \sum_{n=0}^2
d_n \cos n\theta  ,
\label{eqn:XSlinear_Phi_i}
\eea
thus providing 3 more bilinear relations (generally
$c_n$ and $d_n$ are different).

The bilinear relations provide a system of quadratic equations for the multipoles. 
In reality, the angular coverage and precision of the data do not allow for unique
solution of these equations, at least not at the present time. 
Fortunately, as discussed in what follows,
the sum rules for the forward CS provide 
accurate  linear relations, which simplify things a lot.

\subsection{Linear relations:  Sum rules}
The general properties of forward CS, derived from unitarity, causality
and crossing \cite{GellMann:1954db}, allow for it to be expressed
entirely in terms of integrals of total photoabsorption cross sections.
In case of a spin-1/2 target such as the proton, the forward CS is characterized by two scalar
amplitudes, $f(\nu)$ and $g(\nu)$, functions
of the invariant $\nu =(s-u)/4M$, which in the forward kinematics
is identical to the photon lab energy $E_\gamma$. 
The helicity amplitudes are written in terms of the scalar amplitudes as follows:
\begin{equation}
T_{\sigma'\lambda'\sigma\lambda} \stackrel{t=0}{=} 
\chi^{\dagger}_{\la'} \left\{
f(\nu) \, \vec\varepsilon_{\sigma'}^{*} \cdot \vec\varepsilon_{\sigma} +  g(\nu) \,i \, (\vec\varepsilon_{\sigma'}^{*} \! \times \vec\varepsilon_{\sigma} )\cdot\vec\sigma
\right\} \chi_{\la}\,,
\end{equation}
where $\vec\varepsilon_\sigma$ and $\chi_\lambda$ are the photon polarization vector
and the nucleon spinor, with the subscripts showing the corresponding helicities. These forward amplitudes are given by the sum rules
on one hand and by the multipole expansion on the other:
\begin{widetext}
\begin{subequations}
\bea
\eqlab{A_1_A_3_multipoles}
f(\nu) &=& -\,\frac{\alpha}{M} + \frac{\nu^2}{4\pi^2}\int_0^\infty\! \frac{\dd \nu'}{\nu^{\prime \, 2}-\nu^2-i0^+}\, \big[\sigma^{\mathrm{abs}}_{1/2}(\nu')+\sigma^{\mathrm{abs}}_{3/2}(\nu') \big]\nn\\
&=&\frac{\sqrt{s}}{2 M} \sum_{L=0}^\infty (L+1)^2 \left\{ \left(L+2\right) \big(f_{EE}^{(L+1)-}+f_{MM}^{(L+1)-}\big) + L \left(f_{EE}^{L+}+f_{MM}^{L+}\right) \right\} \nn \\
&\stackrel{J< 5/2}{=}& \frac{\sqrt{s}}{M}\left(f_{EE}^{1-}+2 f_{EE}^{1+}+f_{MM}^{1-}+2 f_{MM}^{1+} +6 f_{EE}^{2-}+ 9 f_{EE}^{2+} + 6 f_{MM}^{2-}+9 f_{MM}^{2+} \right),  \\
g(\nu)&= & - \frac{\al\varkappa^2 \nu}{2M^2} + \frac{\nu^3}{4\pi^2}\int_0^\infty\! \frac{\dd \nu'}{\nu'}\, \frac{\sigma^{\mathrm{abs}}_{1/2}(\nu')-\sigma^{\mathrm{abs}}_{3/2}(\nu')}{\nu^{\prime \, 2}-\nu^2-i0^+}\nn\\
&=& \frac{\sqrt{s}}{2 M} \sum_{L=0}^\infty (L+1) \Big\{ \left(L+2\right) \big(f_{EE}^{(L+1)-}+f_{MM}^{(L+1)-}\big) - L \left(f_{EE}^{L+}+f_{MM}^{L+}\right)  -2 L \left(L+2\right) \left(f_{EM}^{L+}+f_{ME}^{L+}\right) \Big\}\nn\\
&\stackrel{J< 5/2}{=}&\frac{\sqrt{s}}{M} \left(f_{EE}^{1-}-f_{EE}^{1+}-6
   f_{EM}^{1+}-6
   f_{ME}^{1+}+f_{MM}^{1-}-f_{MM}^{1+}+ 3 f_{EE}^{2-} - 3 f_{EE}^{2+}+3 f_{MM}^{2-} - 3 f_{MM}^{2+} \right).
\eea
\end{subequations}
\end{widetext}
where $\sigma^\text{abs}_\Lambda$ is the photoabsorption cross section
corresponding to the total helicity $\Lambda$ of the initial $\gamma p$ state, $0^+$
is an infinitesimal positive number. The summation of the multipoles
is done over $L=J-\nicefrac12 = (l\mp \nicefrac12)-\nicefrac12$. Note also that the first term in the sum-rule expressions (due to the proton charge
and anomalous magnetic moment $\varkappa$) are precisely
the Born contributions, whereas the integrals yield the non-Born contributions.

The empirical evaluation of the forward
amplitudes $f(\nu)$ and $g(\nu)$ for proton CS has recently been performed
in~\cite{Gryniuk:2015eza} and \cite{Gryniuk:2016gnm}, respectively.
Thus, the sum rules provide two {\it linear} relations on the multipole
amplitudes. We use these relations to eliminate the residual functions $f_2^R(E_\gamma)$ and $f_3^R(E_\gamma)$ in \Eqref{nonBorn_multipoles_expansion}.

The low-energy expansion of the integrals in \Eqref{A_1_A_3_multipoles}
yield sum rules for the forward combinations of static 
polarizabilities, such as the Baldin sum rule \cite{Baldin} 
(see, e.g., \cite[Sec.\ 5]{Hagelstein:2015egb} for more details). 
Based on the empirical evaluation of these sum rules, we use~\cite{Gryniuk:2015eza,Gryniuk:2016gnm}:
\begin{subequations}
\eqlab{numSR}
\bea 
\alpha_{E1}+\beta_{M1}&=&(14.0 \pm 0.2) \times 10^{-4}\, \mathrm{fm}^3\,,\\
\gamma_0 &\equiv &-\gamma_{E1E1}-\gamma_{M1M1}-\gamma_{E1M2}-\gamma_{M1E2}\nn\\
&=& ( - 0.929 \pm 0.044) \times 10^{-4}\, \mathrm{fm}^4
\eea 
\end{subequations}
to eliminate two out of six global parameters in \Eqref{nonBorn_multipoles_expansion}, our choice being
$\alpha_{E1}+\beta_{M1}$ (so that $\alpha_{E1}-\beta_{M1}$
is fitted) and $\gamma_{M1M1}$.
We thus are left with
four global parameters and four energy-dependent functions.

\setlength{\abovedisplayskip}{7pt plus 3pt minus 3pt}
\setlength{\belowdisplayskip}{7pt plus 3pt minus 3pt}

\section{CS database and fitting strategy}
\label{sec:fitting}
The world database on the unpolarized angular distribution of proton CS, below the pion-production threshold, is summarized in Table~\ref{tab:unpolarized Compton-scattering experiments}, cf.~\cite{Baranov:2001jv,Griesshammer:2012we}.
The number of data points contributed by each experiment is indicated in the column $\text{N}_{\text{data}}$. 
The database is split into $\text{N}_{\mathrm{bins}}=11$ energy bins, with the central values at\footnote{We have tried to optimize the number of
energy-bins to minimize the number of fitting parameters. Thus, we
omitted the data from the very low-energy region (below 50 MeV), such
as those of  Federspiel et al.~\cite{Federspiel:1991yd}, which would
have  relatively low number of points per bin. We have also omitted two 
data points from the same source taken at $65.8$~MeV,
in order to avoid having a separate bin.}
\beq
\label{eq:list_energies}
59,\, 69,\, 79,\, 89, \, 99,\, 109,\, 117,\, 127,\, 135,\, 143,\,  150\, \, \text{MeV}.
\eeq
We fit all these data simultaneously, hence the number of parameters is
$4 + 4 \, \text{N}_{\text{bins}} = 48$.
This is quite a large number, and we perform the fitting in 
two stages: 1) a Monte-Carlo swipe fitting both the static polarizabilities and the residual functions, 
by finding the least $\chi^2$ for a large ensemble of parameters taken from the normal (Gaussian) distribution\footnote{In the interest of full disclosure, the parameters of the normal distribution used in all of our fits are as follows. The 
residual functions are centered at zero
with the width for $f^R_{1,4}(E_\gamma)$  given by $10\times 10^{-4}~\text{fm}^3$, and the width 
for $f^R_{5,6}(E_\gamma)$ given by $10^{-4}\, \text{fm}^4$. 
These choices
for the widths are motivated by 
the ``natural size'' argument based on the known values of
the static polarizabilities, even though, the use of normal distribution
allows for these functions to take any values in principle.
The parameters of the normal distributions
for $\alpha_{E1}-\beta_{M1}$, $\gamma_{E1E1}$, $\gamma_{M1E2}$,
and $\gamma_{E1M2}$ are taken as $\{9.5,-3.3,1.1,0.2\}$ 
for the mean, and $\{2.0,2.0,2.0,2.0\}$ for the width, in units
of $10^{-4}\{\text{fm}^3,\text{fm}^4,\text{fm}^4,\text{fm}^4\}$,
covering the range of the different extractions.}; 2) the $\chi^2$ is further minimized by varying the static polarizabilities
using a standard minimization routine, whilst the residual functions are kept fixed to the values determined in the 
Monte-Carlo swipe.

Our fit to the database of Table~\ref{tab:unpolarized Compton-scattering experiments} results in Fit~0 of Table~\ref{tab:polarizabilities_result}. The results of Fit~1 correspond to the fit where the small (according to 
many existing analyses,
cf.\ the last 3 rows of Table~\ref{tab:polarizabilities_result}) spin polarizability $\gamma_{E1M2}$
is set to zero. The results of the two fits are consistent with each other,
albeit Fit~1 provides a much better accuracy. We take it as a sign of insufficient data quality for the accurate determination of the small
value of $\gamma_{E1M2}$, and keep $\gamma_{E1M2}=0$ in our subsequent 
fits.

\begin{figure}[htb]
\centerline{\includegraphics[width=0.9\linewidth]{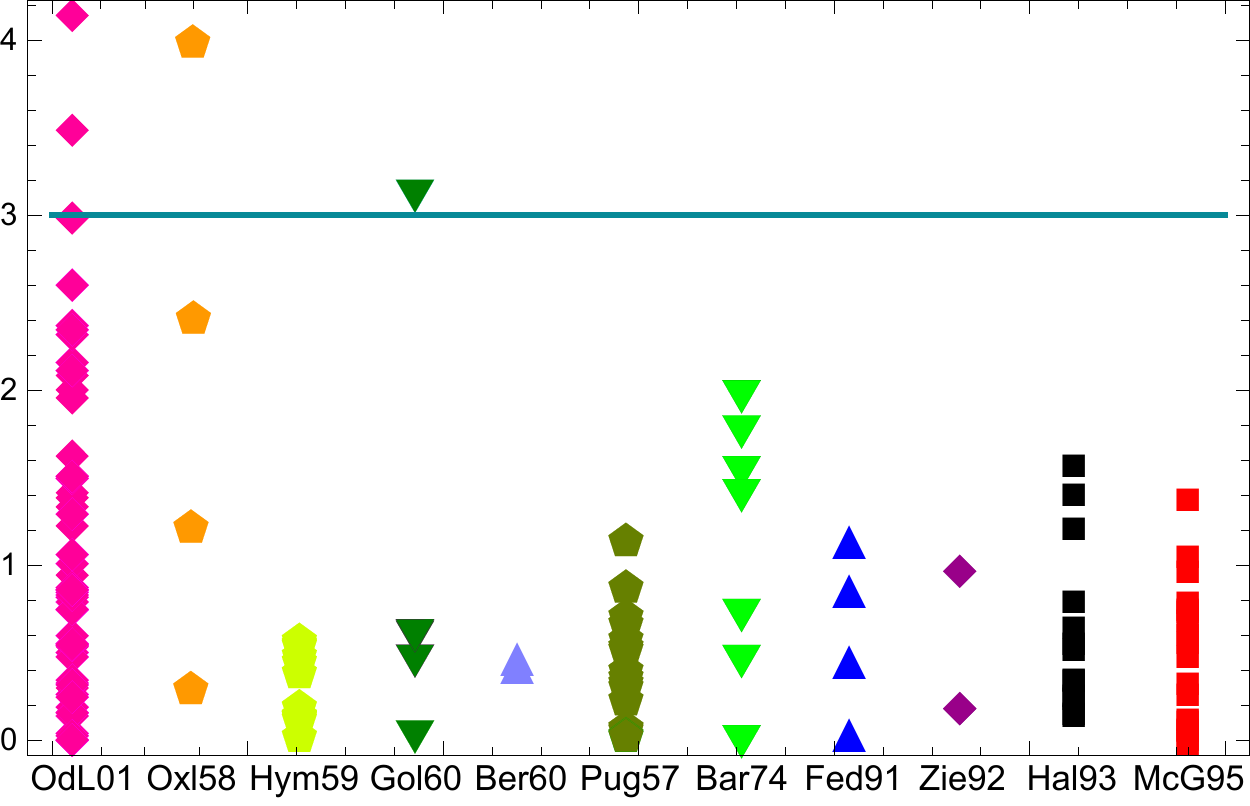}}
\caption{Distribution of $\chi^2$ contributions per data point in Fit 1.
The 4 points above the 3$\si$ line correspond with: Ref.~\cite{OlmosdeLeon:2001zn} at $\{E_\gamma,\vartheta\} =\{ 89$~MeV, $155^\circ\}$ and $\{ 109$~MeV,
$133^\circ\}$; Ref.~\cite{Oxley:1958zz} at $\{60$ MeV, $150^\circ\}$; 
Ref.~\cite{Goldansky:1960zz} at $\{55$ MeV, $150^\circ\}$.}
\label{fig:outliers}
\end{figure}

\begin{figure}[h]
\centerline{\includegraphics[width=0.8\linewidth]{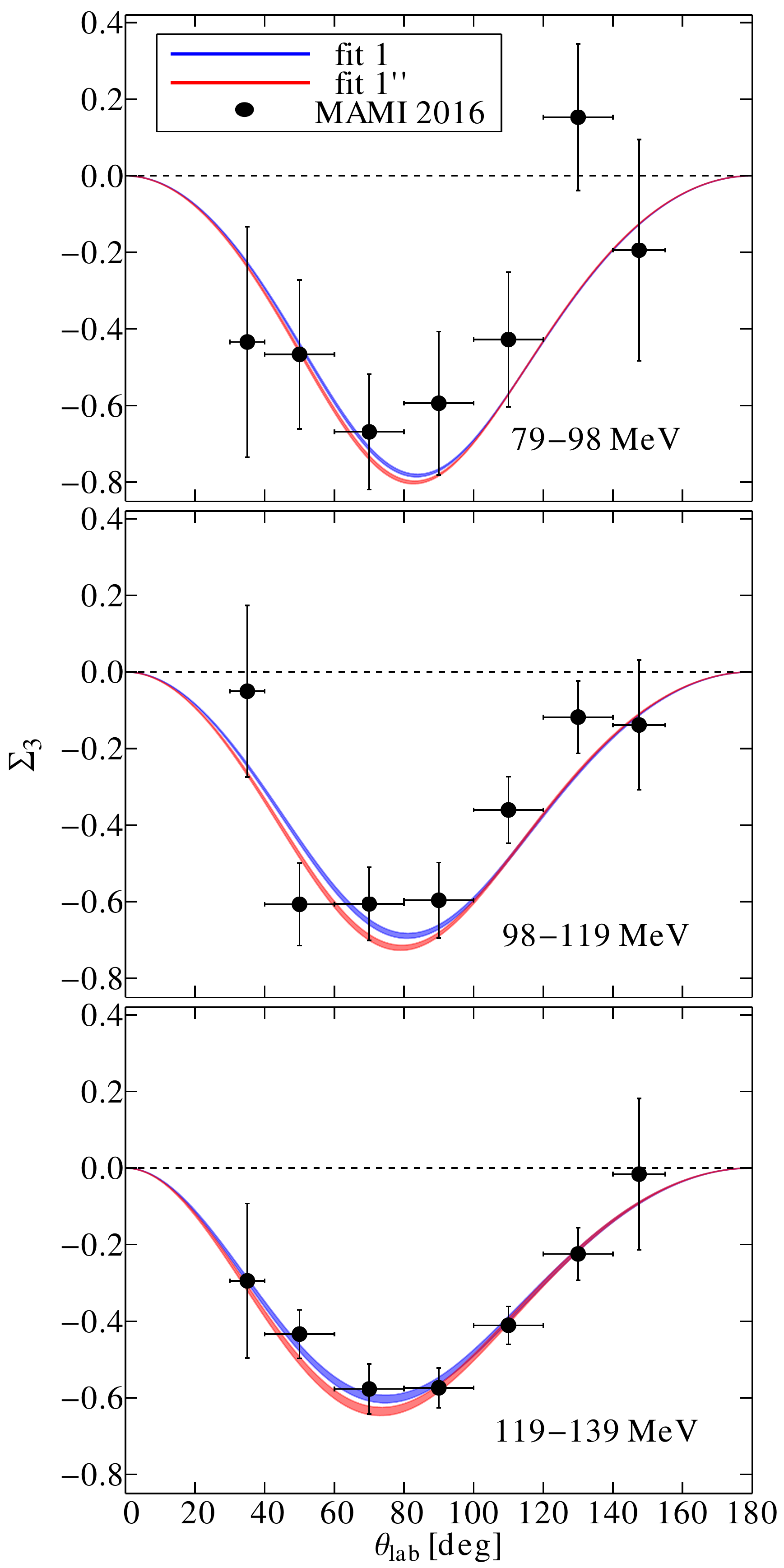}}
\caption{The beam asymmetry as function of the scattering angle at $79-98,\, 98-119$ and  $119-139$~MeV of beam energy. The experimental data 
are from Sokhoyan et al.~\cite{Sokhoyan:2016yrc}. The blue and red bands correspond to the solutions of Fit~$1$ and Fit~$1^{\prime\prime}$, respectively.} 
\figlab{sigma3_fit_MEX}
\end{figure}

\begingroup
     \squeezetable
\begin{table*}[htp]
\caption{Unpolarized proton CS experiments
below the pion-production threshold. The column $\text{N}_{\text{data}}$ indicates the number of data points we use for the fitting. The photon energy, $E_\gamma$, and the scattering angle, $\vartheta$, 
are given in the lab frame. 
}
\label{tab:unpolarized Compton-scattering experiments}
\begin{center}
\begin{tabular}{|L{3.5cm}|C{2cm}|C{2.5cm}|C{2.5cm}|C{2.5cm}|C{2cm}|}
\hline
\rule{0pt}{3.5ex} 
Author& Ref. & $E_{\gamma}$ [MeV]  & $\vartheta$ [deg]  &$\text{N}_{\text{data}}$& Symbol \\
\hline
\rule{0pt}{3.5ex}  
Oxley et al. & \cite{Oxley:1958zz} & 60 & 70-150 & 4 & 
$\rotatebox[origin=c]{90}{\large{\mypentagon{myorange}}}$
\\
\rule{0pt}{3.ex} 
Hyman et al. & \cite{Hyman:1959zz} & 60-128 & 50, 90 &12 &
$\rotatebox[origin=c]{90}{\large{\mypentagon{lightolive}}}$
\\
\rule{0pt}{3.ex} 
Goldansky et al. & \cite{Goldansky:1960zz} & 55 & 75-150 &5 & \vspace*{-0.5em}\rotatebox[origin=c]{180}{\large\mytriangle{mydgreen}}\\
\rule{0pt}{3.ex} 
Bernardini et al. & \cite{Bernardini:1960wya} & 120, 139 & 133 & 2  &
\hspace*{-1.ex}{\large\mytriangle{mylblue}} \\
\rule{0pt}{3.ex} 
Pugh et al.& \cite{Pugh:1957zz} & 59-135 & 50, 90, 135 & 16 & 
$\rotatebox[origin=c]{90}{\large{\mypentagon{olive}}}$
\\
\rule{0pt}{3.ex}
Baranov et al.& \cite{Baranov:1974ec} & 79, 89, 109 & 90, 150 &7 & 
\vspace*{-0.5em}\rotatebox[origin=c]{180}{\large\mytriangle{green}}\\
\rule{0pt}{3.ex}
Federspiel et al.& \cite{Federspiel:1991yd} & 59, 70 &60, 135 & 4 &
\hspace*{-1.ex}{\large\mytriangle{blue}}\\
\rule{0pt}{3.ex}
Zieger et al. & \cite{Zieger:1992jq} & 98, 132 & 180 & 2 &
\rotatebox[origin=c]{45}{\color[HTML]{990089} $\blacksquare$}\\
\rule{0pt}{3.ex}
Hallin et al. & \cite{Hallin:1993ft} & 130-150 &45, 60, 82, 135 & 13 & $\blacksquare $ \\
\rule{0pt}{3.ex}
MacGibbon et al.& \cite{MacGibbon:1995in} & 73-145 & 90-135 & 18 &{\color{red} $\blacksquare $} \\
\rule{0pt}{3.ex}
Olmos de Le{\'o}n et al.& \cite{OlmosdeLeon:2001zn} & 59-149 & 59-155 & 55&
\rotatebox[origin=c]{45}{\color[HTML]{FF0099} $\blacksquare$} \\
 \hline
\end{tabular}
\end{center}
\label{default}
\end{table*}%
\endgroup

\begingroup
     \squeezetable
\begin{table*}[htp]
\caption{The proton scalar and spin polarizabilities in units $10^{-4}\, \text{fm}^3$\,(scalar) and $10^{-4}\, \text{fm}^4$\,(spin), obtained 
in the various fits described in the text, compared with 
the B$\chi$PT predictions~\cite{Lensky:2015awa}, DR calculations~\cite{OlmosdeLeon:2001zn, Babusci:1998ww} (note that only $\alpha_{E1}+\beta_{M1}$ is calculated in DR, with their
difference fitted to CS data), and
an experimental extraction of spin polarizabilities at MAMI~\cite{Martel:2014pba}
(performed using subtracted DRs~\cite{Pasquini:2007hf}).
}
\label{tab:polarizabilities_result}
\begin{center}
\begin{tabular}{L{1.8cm} C{2cm} C{2.cm} C{2.cm} C{2.cm} C{2.cm} C{2.cm} C{1.3cm} }
\hline
\rule{0pt}{3.5ex} 
Source &$\alpha_{E1}$& $\beta_{M1}$ & $\gamma_{E1E1}$ &$\gamma_{M1M1}$&$\gamma_{E1M2}$& $\gamma_{M1E2}$&$\chi^2/\text{point}$ \\
\hline
\rule{0pt}{3.5ex}
Fit $0$  & $12.2 \pm 0.3$ &$1.8 \mp 0.3$ &$-1.6 \pm 2.6$ & $1.8 \pm 1.1$ & $-1.3 \pm 3.7$ & $2.0 \pm 0.7$& $1.35$ \\ 
\hline
\rule{0pt}{3.5ex} 
Fit $1$ & $12.2 \pm 0.3$ &$1.8 \mp 0.3$ &$-3.1 \pm 0.7$ & $1.6 \pm 0.3$ & ${\bf{0.0}}$ & $2.5 \pm 0.7$& $1.35$\\
\rule{0pt}{3.5ex}
Fit $1_{3\sigma}$ & $11.8 \pm 0.3$ &$2.2 \mp 0.3$ &$-2.7 \pm 0.6$ & $1.5\pm 0.3$ & ${\bf{0.0}}$ & $2.2 \pm 0.7$ & $0.97$ \\ 
\rule{0pt}{3.5ex}
Fit $1^\prime$   & $10.6 \pm 0.3$ &$3.4 \mp 0.3$ &$-1.0 \pm 0.8$ & $1.0 \pm 0.3$ & ${\bf{0.0}}$ & $1.0 \pm 0.7$ & $0.99$ \\ %
\rule{0pt}{3.5ex}
Fit $1^{\prime\prime}$ & $10.2 \pm 0.4$ &$3.8 \mp 0.4$ &$-1.2 \pm 0.8$ & $0.6 \pm 0.3$ & ${\bf{0.0}}$ & $1.6 \pm 0.8$ & $ 0.62$ \\ 
\hline
\rule{0pt}{3.5ex}
no $l=2$ & & & & & & &  \\
\rule{0pt}{3.5ex}
Fit $2$ & $11.7 \pm 0.3$ &$2.3 \mp 0.3$ &$-2.6 \pm 0.6$ & $1.1 \pm 0.3$ & ${\bf{0.0}}$ & $2.4 \pm 0.7$ & $ 1.35$ \\ 
\rule{0pt}{3.5ex}
Fit $2^{\prime\prime}$
                  & $10.8 \pm 0.4$ &$3.2 \mp 0.4$ &$-1.9 \pm 0.8$ & $0.7 \pm 0.3$ & ${\bf{0.0}}$ & $2.2 \pm 0.8$ & $ 0.69$ \\ 
 \hline
\rule{0pt}{3.5ex} 
B$\chi$PT& $11.2 \pm 0.7$ & $3.9 \pm 0.7$ & $-3.3 \pm 0.8$ & $2.9 \pm 1.5$ & $0.2 \pm 0.2$ & $1.1 \pm 0.3$ \\ 
\rule{0pt}{3.5ex} 
DR &$12.1$ & $1.6$ & $-3.4$ & $2.7$ & $0.3$ & $1.9$\\
\hline
\rule{0pt}{3.5ex} 
MAMI 2015 & & &$ -3.5 \pm 1.2$ & $3.16 \pm 0.85$ & $-0.7 \pm 1.2$ & $1.99 \pm 0.29$\\
\hline
\rule{0pt}{3.5ex}  
\end{tabular}
\end{center}
\label{results}
\end{table*}%
\endgroup

The three subsequent fits in Table~\ref{tab:polarizabilities_result}
are done upon various ``refinements'' of the database involving deletion
of ``outliers". Namely,
\begin{itemize}
\item in Fit $1_{3\si}$, the outliers are identified 
according to the simple $3\sigma$ rule~\cite{outlier_detection,anomaly_detection}, i.e.,
as those that deviate more than 3$\si$ from Fit 1, see 
\Figref{outliers}.
\item In Fit $1'$, the 4 deleted outliers are: Ref.~\cite{OlmosdeLeon:2001zn} at $\{E_\gamma,\vartheta\} =\{ 89$~MeV, $133^\circ\}$ and $\{ 109$~MeV,
$133^\circ\}$; Ref.~\cite{Oxley:1958zz} at $\{60$ MeV, $120^\circ\}$
and $\{60$ MeV, $150^\circ\}$. Hence, two of the deleted points
are the same as in the previous fit and two are different. 
The latter two point are selected by hand such as to drive the fit
closer to the B$\chi$PT-predicted cross section.

\item In Fit $1''$, we purge the database in accordance to what is
done in $\chi$PT fits as described in~\cite{Griesshammer:2012we}, i.e.:
omit the data of Oxley et al.~\cite{Oxley:1958zz} entirely,
Bernardini et al.~\cite{Bernardini:1960wya} entirely, Baranov et al.~\cite{Baranov:1974ec} at $\theta_\mathrm{lab}=150^\circ$,
and Olmos de Le\'on et al.~\cite{OlmosdeLeon:2001zn} at \{$109$~MeV, $133^\circ$\}. Furthermore, as in~\cite{Griesshammer:2012we}, 
we add $5\%$ systematic uncertainty (point-to-point in quadrature with
the statistical error) to the points of Ref.~\cite{OlmosdeLeon:2001zn}.
Unlike \cite{Griesshammer:2012we}, we do not include the floating normalization factors. Also, the points of Federspiel et al.~\cite{Federspiel:1991yd} are treated as described in the footnote (i.e., omitting
them below 50 and at 65.8 MeV), rather than just removing the point
\{$44$~MeV, $135^\circ$\} as done in~\cite{Griesshammer:2012we}.
\end{itemize}

 We do not include the data on beam asymmetry
 in our fits, since the only data (below pion-production threshold),
 coming from the pilot experiment at MAMI~\cite{Sokhoyan:2016yrc},
 are of significantly poorer quality 
 compared to the unpolarized data. Hopefully, the
 currently running A2/MAMI experiment will improve the accuracy
 for this observable, and thus play a crucial role in 
 an accurate determination the magnetic polarizability $\beta_{M1}$~\cite{Krupina:2013dya}. At present we only verify
 that all our fits are in agreement with the pilot data~\cite{Sokhoyan:2016yrc}, see~\Figref{sigma3_fit_MEX}.

\section{Results and discussion}
\label{sec:results}

\begin{figure*}[htb]
\centerline{
\includegraphics[width=0.7\linewidth]{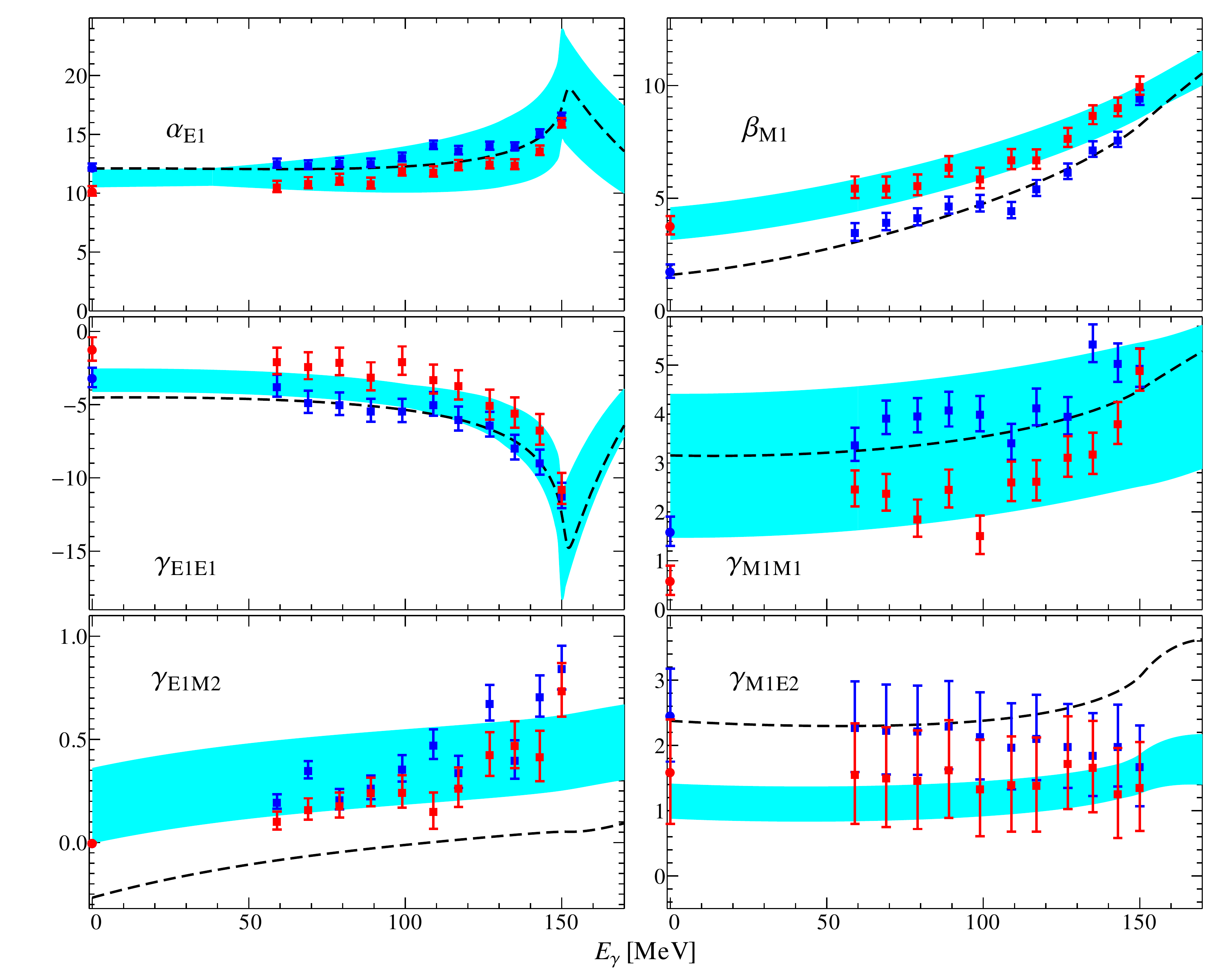}}
\caption{The dynamical polarizabilities as functions of the photon lab energy $E_\gamma$. The black dashed curve is the subtracted DR result~\cite{Hildebrandt:2003fm}, the cyan band corresponds to the B$\chi$PT prediction, and the points with error bars are the results of fit~$1$ (blue) and fit~$1^{\prime \prime}$ (red).} 
\label{Dynamical_pols}
\end{figure*}

Table~\ref{tab:polarizabilities_result} presents the static polarizability values resulting from the 5 fits described in the previous section.
The last column shows $\chi^2/\text{point}$, a measure of the fit quality. 
The obtained polarizabilities can be compared with the last 3 rows showing respectively 
the B$\chi$PT prediction, DR extraction, and the first experimental extraction of the spin polarizabilities (MAMI 2015). 

The striking result here 
is that the polarizability values are fairly sensitive
to the slight refinements of the database. For example, for $\beta_{M1}$
we obtain the values ranging from 1.8(3) using the original database
in Fit~1 to 3.8(4) using an improved 
one in Fit~$1^{\prime\prime}$. The latter modification of the
database is similar to the one used in the $\chi$PT fits
of McGovern et al.~\cite{Griesshammer:2012we,McGovern:2012ew,Lensky:2014efa}, which could explain why the $\chi$PT fits are significantly different from the
DR fits, which in particular yield a low value of $\beta_{M1}$.

Let us emphasize that the B$\chi$PT row in the Table
is not an extraction from CS data, but is rather a prediction,
albeit of a low order~\cite{Lensky:2009uv,Lensky:2015awa}. Nonetheless,
if we are to take the claimed uncertainties seriously, we 
must conclude that the refined databases agree
somewhat better with $\chi$PT.

Besides the static polarizabilities, our fits yield the
multipole amplitudes at the considered energy bins. 
The non-Born multipoles can equivalently be represented by the so-called
{\it dynamical polarizabilities} (see, e.g., \cite[Sec.\ 2]{Hagelstein:2015egb}
for definition). The blue (red) points in 
Fig.~\ref{Dynamical_pols} show the dynamical polarizabilities
resulting from Fit 1 ($1^{\prime\prime}$). Note that the point at zero energy corresponds with the static
polarizability.
The error bars result from the uncertainties on the fit 
parameters. The results are
compared with the B$\chi$PT 
(cyan bands) and DR (dashed lines)
results.  Again, we see that our solution based on the raw database
(Fit 1) agrees well with DR calculation, whereas the one based on the refined database (Fit $1''$) agrees better with B$\chi$PT. 

Therefore, the differences between the $\chi$PT and DR results for polarizabilities are likely to be caused by deficiencies in the
experimental database. How to resolve those? We first of all need
to find the place where the differences among the different
fits are largest. For the unpolarized 
cross section, the ``sweet spot'' is apparently at $E_\gamma \simeq 109$
MeV and backward angles, see \Figref{Unpol_En59_En109}. 
At both higher and lower energies
the difference among the fits quickly diminishes, cf.~\Figref{UnpolXS}.
Hence the best hope for resolving this ``database consistency problem''
is to obtain new precise cross-section data at energies close to 109 MeV.

\begin{figure}[h]
\centerline{\includegraphics[width=0.9\linewidth]{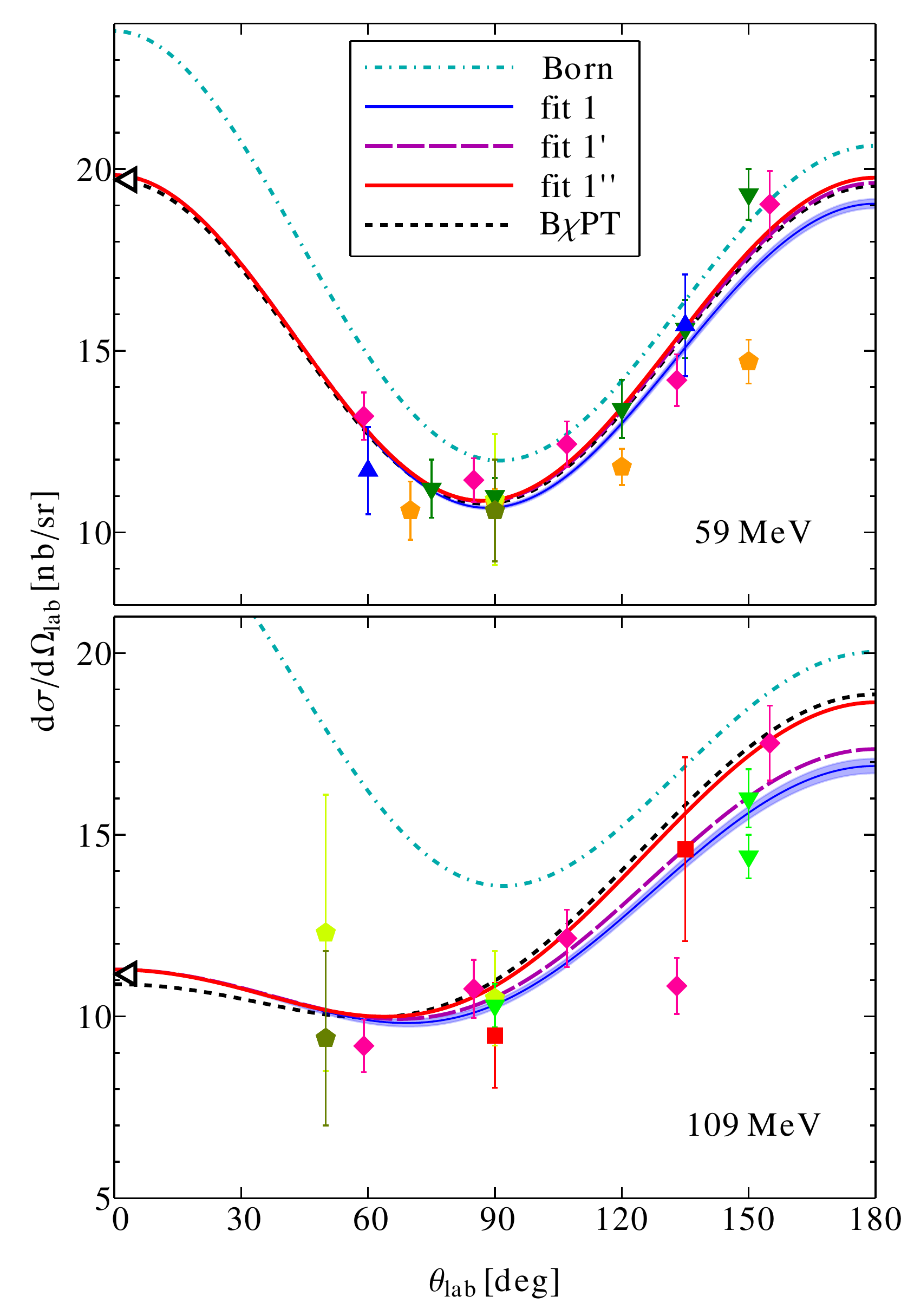}}
\caption{Unpolarized cross section of proton CS as function of 
scattering angle in the lab frame at photon-beam energy $59$~MeV (top panel) and $109$~MeV (bottom panel). The legend for experimental data points is given in Table~\ref{tab:unpolarized Compton-scattering experiments}. The error band
on the fit 1 is obtained by the simple error-propagation of the
fit values of the static polarizabilities only.
The other fits have a comparable error band, which is not shown
here for clarity. 
 } 
\figlab{Unpol_En59_En109}
\end{figure}

\begin{figure}[h]
\centerline{\includegraphics[width=1.03\linewidth]{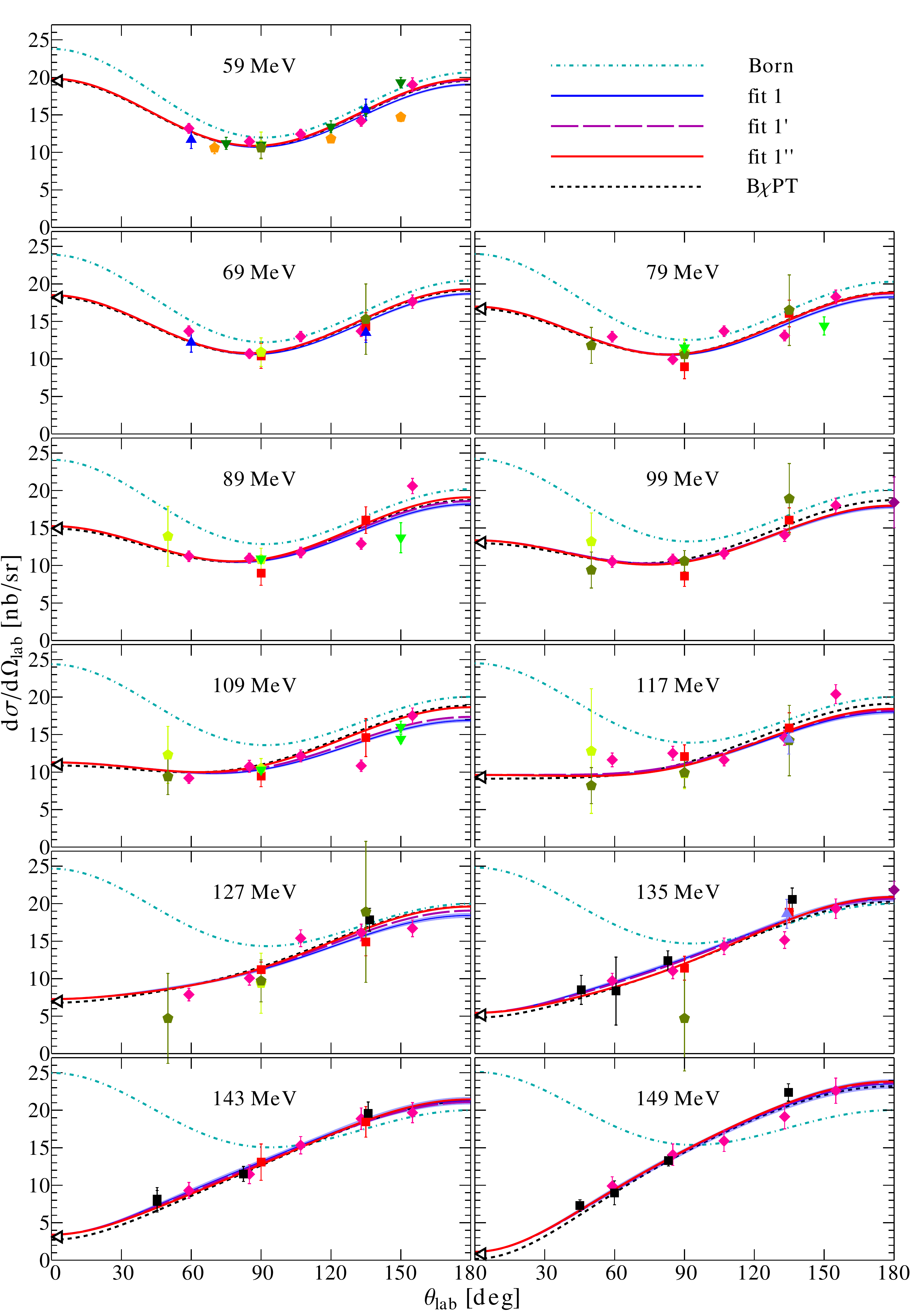}}
\caption{Same as in \Figref{Unpol_En59_En109}, for all the energy bins.} 
\figlab{UnpolXS}
\end{figure}

Let us consider this energy region in more detail. In \Figref{Unpol_En59_En109}, besides the data and the results of 3 fits,
we show the Born contribution (dash-dotted curve)
and the B$\chi$PT prediction~\cite{Lensky:2009uv,Lensky:2015awa} (dotted curve). 
The deviation from the Born contribution is the effect
of (dynamical) polarizabilities we are after. The polarizability
contribution is at low-energy dominated by the scalar dipole polarizabilities, $\al_{E1}$ and $\beta_{M1}$, but already at 
109 MeV the spin polarizabilities start to play a crucial role.

To see this, consider Table~\ref{interplay}, where the forward
and backward combinations of scalar and spin polarizabilities are presented.
In the fits the forward combinations are fixed by the sum rules, 
\Eqref{numSR}, whereas the backward combinations,
$\al_{E1} -\beta_{M1}$ and $\ga_\pi = -\gamma_{E1E1}+\gamma_{M1M1}-\gamma_{E1M2}+\gamma_{M1E2}$ are different from fit to fit.
Fit 1 has the highest value of $\al_{E1} -\beta_{M1}$ and hence
has the biggest deviation from the Born term at 59 MeV; the $\gamma_\pi$
value is not important at these energies. Fits $1'$ and $1''$ have
$\al_{E1} -\beta_{M1}$ close to B$\chi$PT and as the result the
three curves practically coincide at 59 MeV.

However, at 109 MeV,
Fit~$1^\prime$ converges to Fit~1 precisely 
because of the different $\ga_\pi$ value. The similar effect
for Fit~$1^{\prime\prime}$ is diminished by the difference in the value 
of $\al_{E1} -\beta_{M1}$. Thus, at these energies
the scalar and spin polarizabilities are rather entangled and cannot
be extracted independently from this observable. The present PWA,
on the other hand, provides a basis for a simultaneous extractions
of $\al_{E1} -\beta_{M1}$ and the backward spin polarizability 
$\ga_\pi$.

\begin{table}[htp]
\caption{Results the fits 1, $1^\prime$ and  fit $1^{\prime\prime}$ for the forward and backward
combinations of polarizabilities compared to the corresponding values from the B$\chi$PT~\cite{Lensky:2015awa} and DR~\cite{OlmosdeLeon:2001zn, Babusci:1998ww} calculations.
}
\begin{center}
\begin{tabular}{L{1cm} C{1.5cm} C{1.5cm} C{1.5cm} C{1.5cm}}
\hline
\rule{0pt}{3.5ex} 
 & $\alpha_{E1}+\beta_{M1}$& $\gamma_0$ &  
$\alpha_{E1}-\beta_{M1}$ & $\gamma_\pi$ \\
\hline
\rule{0pt}{3.5ex} 
 Fit 1 & $14.0$ & $-0.93$ &  
$10.5 \pm 0.4$ & $7.2 \pm 1.0$ \\
\rule{0pt}{3.5ex} 
 Fit $1^{\prime}$ & $14.0$ & $-0.93$ &  
$7.2 \pm 0.6$ & $3.0 \pm 1.1$ \\ 
\rule{0pt}{3.5ex} 
 Fit $1^{\prime\prime}$ & $14.0$ & $-0.93$ &  
$6.4 \pm 0.6 $ & $3.5 \pm 1.2$ \\ 
\hline
\rule{0pt}{3.5ex} 
B$\chi$PT & $15.1 \pm 1.0$ & $-0.9 \pm 1.4$ &  
$7.3 \pm 1.0 $ & $7.2 \pm 1.7$ \\
\rule{0pt}{3.5ex} 
DR & $13.7$ & $-1.5$ &  
$10.5 $ & $7.8$ \\
\hline
\rule{0pt}{3.5ex}  
\end{tabular}
\end{center}
\label{interplay}
\end{table}%

\section{Summary and Conclusion}
\label{sec:Conclusions} 

We presented a first partial-wave analysis of proton Compton scattering
data below the pion-production threshold ($E_\gamma \lesssim 150$ MeV).
The only approximations, or model-dependent assumptions, we made 
concern the truncation of the partial-wave expansion:
\begin{itemize}
\item we account for the lowest $l=1$ and 2 terms, neglecting $J\geq 5/2$ contributions;   
\item for the $l=2$ multipoles, we assume the values given
by the NNLO B$\chi$PT calculation \cite{Lensky:2009uv,Lensky:2015awa}, and check that the results
do not change qualitatively if we put them to 0 (cf.\ Fit 2
variety in Table~\ref{tab:polarizabilities_result}).
\end{itemize}
The proper low-energy behavior of the (non-Born piece of) multipoles is ensured through the parameterization in terms of lowest static polarizabilities, see \Eqref{nonBorn_multipoles_expansion}.
The sum rules for the forward amplitudes impose two 
linear relations on the multipoles, leaving us with only four
of the six amplitudes to be determined from the Compton angular-distribution data. The accuracy of the resulting solutions is significantly
improved by setting (the small spin polarizability) $\gamma_{E1M2}=0$
by hand. 

The extracted multipoles depend significantly upon very mild refinements of the world database of proton Compton scattering. The characteristic difference between the state-of-art DR and $\chi$PT analyses is likely to be explained by the database inconsistencies, rather than by differences
in the theoretical framework. We claim that these inconsistencies
are best to be addressed by a new precise measurement of
the angular distribution at $E_{\gamma}\approx 109$ MeV and
backward angles (cf.~\Figref{Unpol_En59_En109}). Accurate
data on polarized observables, such as the beam asymmetry, could be
helpful too. 

The ongoing Compton scattering experiment by the A2 Collaboration at MAMI may soon provide a considerable improvement 
of the database, in both the angular distribution and beam asymmetry.
Until then, the static polarizabilities may continue to be extracted
in a rather wide range of values, manifested by our fit results 
in Table~\ref{results}.

\acknowledgments

This work is supported by the Deutsche Forschungsgemeinschaft (DFG) 
through the Collaborative Research Center [The Low-Energy Frontier of the Standard Model (SFB 1044)]. 


\end{document}